\documentclass{article}
\usepackage{spconf}
\usepackage{amsmath,graphicx}

\usepackage{cite}
\usepackage{amsmath,amssymb,amsfonts}
\usepackage{url}
\usepackage{algorithmic}
\usepackage{textcomp}
\usepackage{xcolor}
\usepackage{tcolorbox}
\usepackage{booktabs} 
\usepackage{amsthm} 
\usepackage{makecell, cellspace}
\usepackage{caption}
\usepackage{physics}
\usepackage{multirow}
\usepackage{hyperref} %
\usepackage{cleveref}
\usepackage{array}
\usepackage{subcaption}
\usepackage[table]{xcolor}
\usepackage{tabularx}
\usepackage{ragged2e}
\definecolor{pastelblue}{HTML}{D4F1F9}
\definecolor{pastelyellow}{HTML}{FFF9C4}
\definecolor{pastelpink}{HTML}{FFD6D6}
\definecolor{pastellavender}{HTML}{E8D5FF}
\definecolor{pastelmint}{HTML}{D5F5E3}
\definecolor{pastelpeach}{HTML}{FFE0CC}
\definecolor{tabblue}{RGB}{31,119,180}
\definecolor{taborange}{RGB}{255,127,14}

\definecolor{tabpurple}{RGB}{148,103,189}

\newcommand{\E}{\mathbb{E}}

\newcolumntype{L}[1]{>{\raggedright\arraybackslash}p{#1}}
\newcolumntype{C}{>{\centering\arraybackslash}X}

\def\BibTeX{{\rm B\kern-.05em{\sc i\kern-.025em b}\kern-.08em
    T\kern-.1667em\lower.7ex\hbox{E}\kern-.125emX}}

\newcommand{\bt}[1]{\mbox{$\bf #1$}}

\newcommand{\img}{\bt x}
\newcommand{\cimg}{\hat{\bt x}}

\DeclareMathOperator*{\argmin}{arg\,min}

\def\thetavec{\pmb{\theta}}

\title{Rate-Distortion Optimization with Non-Reference \\ Metrics for UGC compression 
}
\name{\hspace{-2.35em}Xin Xiong$^\textsc{{\color{magenta}sc}*}$, Samuel Fernández-Menduiña$^\textsc{{\color{magenta}sc}*}$, Eduardo Pavez$^\textsc{\color{magenta}sc}$,  Antonio Ortega$^\textsc{\color{magenta} sc}$, Neil Birkbeck$^\textsc{\color{magenta} g}$, Balu Adsumilli$^\textsc{\color{magenta} g}$ \thanks{This work was funded in part by a YouTube gift. *Equal contribution.}}
\address{$^\textsc{\color{magenta}sc}$University of Southern California, Los Angeles, CA, USA\\ $^\textsc{\color{magenta}g}$Google Inc, Mountain View, CA, USA}

\title{Rate-Distortion Optimization for Ensembles of Non-Reference Metrics}
\begin{document}
\ninept
\maketitle
\begin{abstract}
Non-reference metrics (NRMs) can assess the visual quality of images and videos without a reference, making them well-suited for the evaluation of user-generated content. Nonetheless, rate-distortion optimization (RDO) in video coding is still mainly driven by full-reference metrics, such as the sum of squared errors, which treat the input as an ideal target. A way to incorporate NRMs into RDO is through linearization (LNRM), where the gradient of the NRM with respect to the input guides bit allocation. While this strategy improves the quality predicted by some metrics, we show that it can yield limited gains or  degradations when evaluated with other NRMs. We argue that NRMs are highly non-linear predictors with locally unstable gradients that can compromise the quality of the linearization; furthermore, optimizing a single metric may exploit model-specific biases that do not generalize across quality estimators. Motivated by this observation, we extend the LNRM framework to optimize ensembles of NRMs and, to further improve robustness, we introduce a smoothing-based formulation that stabilizes NRM gradients prior to linearization. Our framework is well-suited to hybrid codecs, and we advocate for its use with overfitted codecs, where it avoids iterative evaluations and backpropagation of neural network–based NRMs, reducing encoder complexity relative to direct NRM optimization. We validate the proposed approach on  AVC and Cool-chic, using the YouTube UGC dataset. Experiments demonstrate consistent bitrate savings across multiple NRMs with no decoder complexity overhead and, for Cool-chic, a substantial reduction in encoding runtime compared to direct NRM optimization.
\end{abstract}
\section{Introduction}
\label{sec:intro}
Non-reference metrics (NRMs) \cite{mittal2013making}, which assess the perceptual quality of images and videos without a pristine reference, are important for the compression of user-generated content (UGC) \cite{wang2019youtube}. UGC is typically noisy due to motion blur, non-ideal exposure, and artifacts from prior compression. 
Once uploaded to platforms like YouTube and TikTok, UGC is re-encoded at different bitrates and resolutions to support adaptive streaming \cite{seufert2014survey}. In such pipelines, where the original content is unreliable, NRMs are popular for evaluating perceptual quality  \cite{pavez2022compression, wang2019youtube}. Despite their widespread adoption for quality evaluation, NRMs are still rarely used as bit-allocation objectives during compression \cite{fernandez2025ratedist}. 
Instead, rate-distortion optimization (RDO) in hybrid (i.e., block-based) \cite{sullivan_rate_1998} and overfitted \cite{ladune2023cool} codecs is mostly driven by full-reference metrics (FRMs), such as the sum of squared errors (SSE) \cite{ortega_rate-distortion_1998}, which assume the input is an ideal reference. Hence, distortion converges to perfect quality (as measured by the FRM) as bitrate increases \cite{fernandez2025ratedist}. While appropriate for pristine content, FRMs encourage the preservation of artifacts when applied to UGC, which leads to suboptimal compression \cite{xiong_rate_2023}.

Thus, UGC setups can benefit from using NRMs in RDO. The strategy to incorporate them depends on the codec architecture. In overfitted codecs \cite{ladune2023cool}, the NRM can be included directly in the loss function and optimized end-to-end via gradient descent, as we show in~\Cref{sec:exper}. However, overfitted codecs require multiple evaluations of the RD cost during encoding. Since modern NRMs are mostly implemented as deep neural networks, their repeated evaluation and backpropagation add substantial encoder complexity and runtime overhead \cite{philippe2025perceptually}.
For hybrid encoders, given that most NRMs map the input to a single value, obtaining per-pixel or per-block importance is not straightforward. As a result, RDO with NRMs for hybrid encoders may require iterative encoding, decoding, and metric evaluation.  
To address this issue, \cite{fernandez2025ratedist} proposed using the gradient of the metric evaluated at the input to guide bit allocation, a strategy termed linearized NRM (LNRM), which drastically reduces the complexity of accounting for NRMs in hybrid codecs.

\begin{figure*}[t]
\centering
\includegraphics[width=\linewidth]{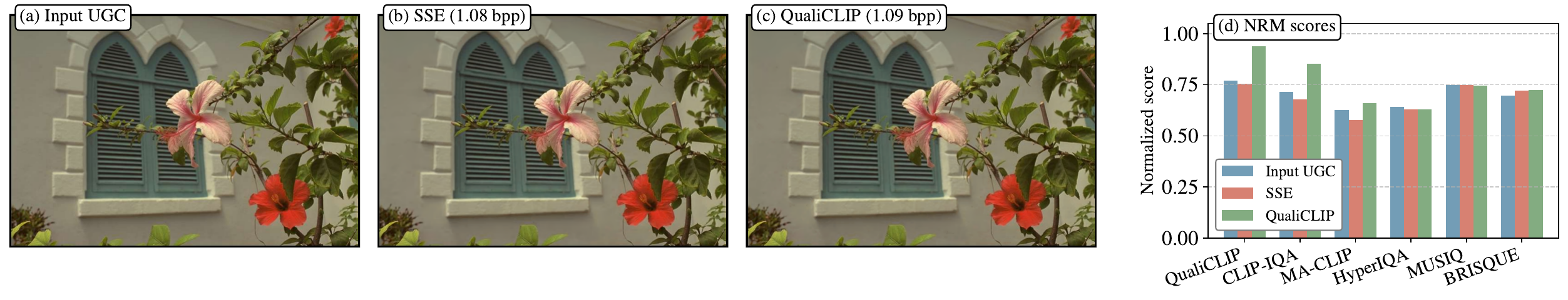}
\caption{(a) Synthetic UGC obtained by compressing a KODAK image \cite{kodak1993kodak} using JPEG \cite{wallace_jpeg_1991} (with $Q=60$).
(b) Cool-chic reconstruction, optimizing for SSE (PSNR: $42.98$ dB).
(c) Cool-chic reconstruction, optimizing for QualiCLIP with SSE regularization (PSNR: $42.63$ dB).
(d) Normalized scores (between $0$ and $1$) reported by multiple NRMs. While optimizing the bit-allocation for QualiCLIP improves the quality as predicted by some metrics, it yields limited improvements or even degradations in others.
}
\label{fig:SSE_vs_Dbcnn_combine}
\end{figure*}

While LNRM makes NRM-based RDO computationally efficient, NRMs are highly non-linear mappings from images to scalars, and their input gradients may exhibit local instability, i.e., the metric’s response can vary sharply even for very small input changes \cite{liu2024defense}. Moreover, NRMs are imperfect quality estimators with model-specific biases, so that different NRMs produce different results for the same input. Experimentally, we observe, with both hybrid and overfitted codecs, that optimizing an LNRM derived from a specific (target) NRM often leads to significant improvements for the target NRM, while we observe marginal improvements (or even degradations) for other NRMs (\Cref{fig:SSE_vs_Dbcnn_combine}).
Although in some settings this might be desirable, gains that are more consistent across an ensemble of NRMs are more likely to reflect perceptual quality improvements.

%

In this work, we address these issues along two directions. First, we extend the LNRM formulation \cite{fernandez2025ratedist} to ensembles of metrics, thereby reducing sensitivity to the choice of NRM and encouraging improvements across multiple, potentially unrelated quality predictors. 
Second, we introduce a smoothing-based optimization strategy, in which for a given NRM we average the scores obtained over small Gaussian perturbations of the input 
before linearization. 
This smoothed gradient strategy, similar to methods developed in the machine learning literature \cite{cohen2019certified, smilkov2017smoothgrad}, computes an average of slightly perturbed gradients, attenuating sharp local variations and favoring directions that remain stable under small input perturbations. 
This improves the reliability of the first-order linearization underlying LNRM. 
Our approach can be incorporated into hybrid codecs, improving performance not only for the target NRMs but also across an ensemble of metrics. We further argue for its use with overfitted codecs, where linearization substantially reduces encoding complexity relative to direct metric optimization. When combined with smoothing, further computational gains are obtained by avoiding iterative evaluations of the smoothed objective.

We validate our framework using AVC \cite{wiegand_overview_2003} and Cool-chic \cite{ladune2023cool} for several NRMs \cite{chen2024topiq, agnolucci2024quality, zhang2020dbcnn}. 
We test frames sampled from the YouTube UGC dataset \cite{wang2019youtube}, setting the RDO to optimize different NRMs using our approach, considering both ensembles of metrics and smoothed objectives. For Cool-chic, we also consider direct NRM optimization. Results in \Cref{sec:exper} show that our method can yield improvements in diverse NRMs, with no decoder complexity overhead and, for Cool-chic, a substantial reduction in runtime relative to optimizing the NRM directly.  

\section{Preliminaries}
\label{sec:prelim}
\noindent\textbf{Notation.} Uppercase and lowercase bold letters, such as $\bt A$ and $\bt a$, denote matrices and vectors, respectively.
The $n$th entry of the vector $\bt a$ is $a_n$. Regular letters denote scalar values.

\subsection{Rate-distortion optimization}
Both hybrid \cite{bross2021overview, wiegand_overview_2003} and overfitted \cite{ladune2023cool} codecs aim to find parameters $\thetavec$ that optimize a rate-distortion (RD) cost. Given $\bt x$, an input with $n_p$ pixels, and $\cimg(\thetavec)$, its compressed version with parameters $\thetavec$, full-reference RDO aims at minimizing:
\begin{equation}
\label{eq:rd_cost_orig}
    J(\cimg(\thetavec)) = d(\cimg(\thetavec), \img) + \lambda \, r(\cimg(\thetavec)),
\end{equation}
where $d(\cdot, \cdot)$ denotes the distortion metric, $r(\cdot)$ the bitrate, and $\lambda$ controls the RD trade-off. 
For the purpose of this paper, the key difference between hybrid and overfitted codecs is the domain of the coding parameters $\thetavec$ and optimization strategy used to minimize \eqref{eq:rd_cost_orig}.

\smallskip

\noindent\textbf{Hybrid codecs.} The parameters  $\thetavec = [\theta_1, \hdots, \theta_{n_b}] \in \mathbb{N}^{n_b}$ take discrete values, which may represent block partitioning, prediction mode, or quantization step. Given a distortion term that splits block-wise $d(\cimg(\thetavec), \img) = \sum_{i = 1}^{n_b} d_i(\cimg_i(\theta_i) \img_i)$, e.g., the sum of squared errors (SSE), where $\bt x_i$ is the $i$th block of the input and $\cimg_i(\theta_i)$ its compressed version with parameters $\theta_i$, we have block level RDO: $\theta_i^\star = \argmin_{\theta_i} \ d_i(\cimg_i(\theta_i), \img_i) + \lambda \, r(\cimg_i(\theta_i))$.

\smallskip

\noindent \textbf{Overfitted codecs.} We focus on Cool-chic \cite{ladune2023cool}. The continuous parameters $\thetavec$, namely,  the latent representation of the image, the entropy coding parameters, and the weights of the synthesis CNN, are optimized via gradient descent. Since quantization and entropy coding are not differentiable \cite{balle_end_2016}, a differentiable approximation of the RD cost in \eqref{eq:rd_cost_orig} is used instead. 
The parameters are updated over thousands of iterations using this cost, which must be evaluated and backpropagated at every iteration. 

\subsection{Linearized NRM}
A method to perform RDO with hybrid codecs using a NRM as distortion metric was proposed in \cite{fernandez2025ratedist}. Let the NRM be $b(\cdot)$. Then,
\begin{equation}
\label{eq:rdcost}
    \thetavec^\star = \argmin_{\thetavec} \  b(\cimg(\thetavec)) + \tau\norm{\img - \cimg}_2^2 + \lambda \, r(\cimg(\thetavec)),
\end{equation}
where $\tau$ controls the SSE-NRM trade-off. When the metric is differentiable and the gradients exist, a first order Taylor expansion to $b(\cimg(\thetavec))$ around the input image $\img$ yields
\begin{equation}
\label{eq:LNRM}
\thetavec^\star = \argmin_{\thetavec} \ \nabla b(\img)^\top(\cimg(\thetavec) - \img)+ \tau\norm{\img - \cimg}_2^2  + \lambda \, r(\cimg(\thetavec)).
\end{equation}
The term  $\nabla b(\img)^\top(\cimg(\thetavec) - \img)$ is called linearized NRM (LNRM). The dependence on the compressed image is linear, which implies that the metric can be evaluated block-wise in hybrid codecs.


\section{Optimization with NRM}
\label{sec:optim}
Since the gradient in \eqref{eq:LNRM} depends only on the input image, the NRM needs to be evaluated (and backpropagated) only once. We leverage this property to optimize efficiently for ensembles of NRMs.
\subsection{Ensembles of metrics}
We consider an ensemble of metrics, $\ell_c(\img) \doteq \sum_{i=1}^{n_m} \, \tau_i \, b_i(\img)$,
where $b_i(\cdot)$ is the $i$th NRM and $\tau_i$ is the corresponding weight. By applying a first-order Taylor expansion around $\img$, we obtain:
\begin{equation}
\label{eq:Combined_NRMs}
\ell_c(\cimg(\thetavec)) = \ell_c(\img) + \nabla \ell_c(\img)^\top (\cimg(\thetavec)-\img) + o(\norm{\cimg(\thetavec)-\img}_2^2),
\end{equation}
where the ensemble gradient satisfies $\nabla \ell_c(\img) = \sum_{i=1}^{n_m} \, \tau_i 
\, \nabla b_i(\img)$. Thus, we can optimize for an ensemble of metrics by 1) evaluating each NRM and then computing the gradient of the metric with respect to the input, and 2) evaluating an LNRM based on the weighted average of the gradients. LNRMs decouple NRM evaluation from the coding loop: we construct and optimize an LNRM for each input and ensemble of NRMs. As a result, evaluating the LNRM during encoding is inexpensive once the gradients are available. 

\smallskip

\begin{figure}[thbp]
    \centering
    \includegraphics[width=\linewidth]{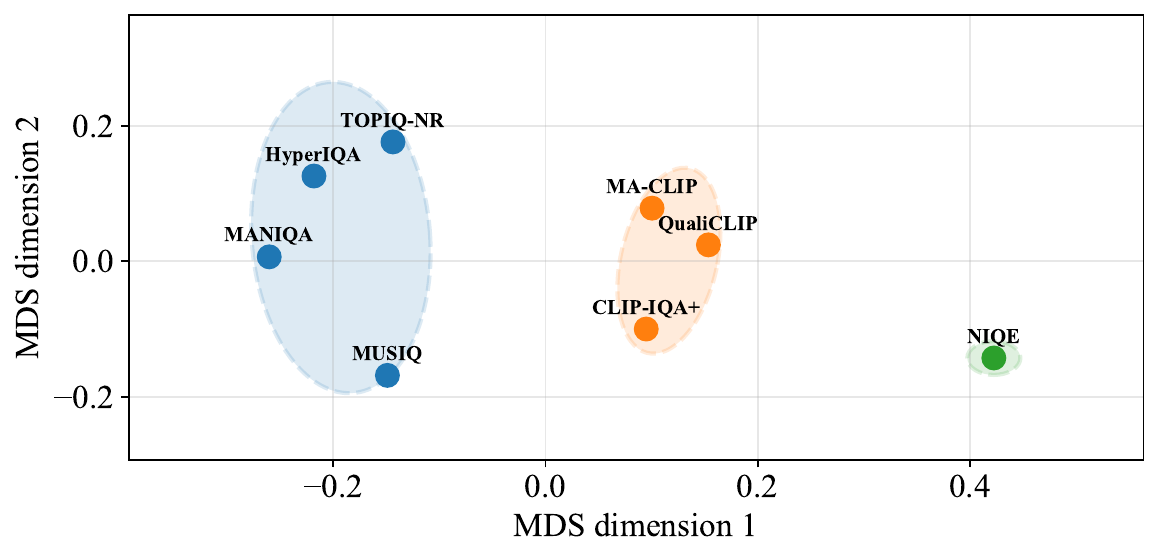}
    \caption{
    \label{fig:metrics_corr}
    Multidimensional scaling (MDS) \cite{kruskal1964multidimensional} for NRM correlation. We show $\bt d_{i, j} = 1 - | \bt p_{i, j}|$, with $\bt p_{i, j}$ the 2D projection of the rank correlation between NRMs $i$ and $j$. We identify three clusters.}
\end{figure}

\noindent\textbf{Ensemble composition.} Since different NRMs may capture different and weakly correlated properties of the image, one way to construct the ensemble is to complement these behaviors rather than reinforce a single notion of quality. Thus, the ensemble will promote improvements that are consistent across diverse and partially independent NRMs. To this end, we analyze the correlation between the scores produced by different NRMs over a set of images. Given a dataset of compressed reconstructions at various bitrates, we compute pairwise correlations between metric outputs. Metrics with high correlation tend to respond similarly to changes in compression parameters and may provide redundant guidance during optimization. Conversely, metrics with low correlation are more likely to capture complementary properties. We show an example in \Cref{fig:metrics_corr} and provide guidance on how to choose the weights $\tau_i$ in \Cref{sec:regul}.

\begin{figure}[t]
    \centering
    
    \begin{subfigure}{\linewidth}
        \centering
        \includegraphics[width=\linewidth]{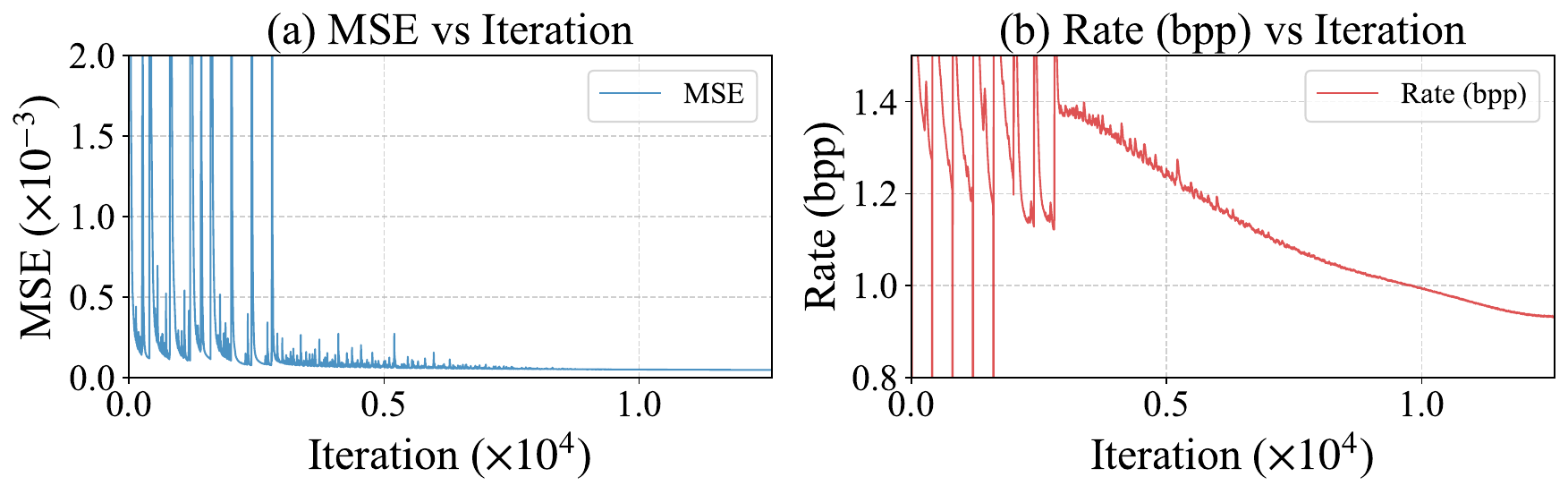}
    \end{subfigure}
    
    \begin{subfigure}{\linewidth}
        \centering
        \includegraphics[width=\linewidth]{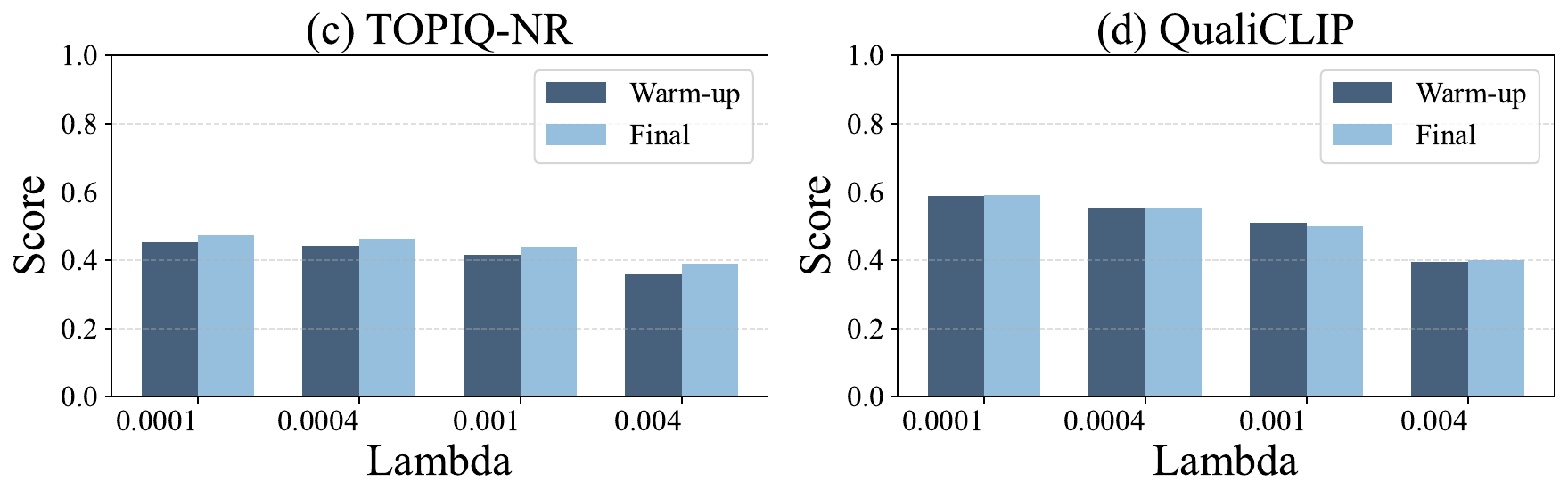}
    \end{subfigure}
    
    \caption{(a) MSE and (b) rate evolution across iterations for Cool-chic in one image. While the MSE converges fast, the decrease in rate is more gradual. (c-d) Average NRM score (50 images), showing warm-up reconstructions achieve near-final scores, justifying \eqref{eq:calibration}.}
    \label{fig:warmup}
\end{figure}

\subsection{Smoothed LNRM}
An approach to improve the robustness of the LNRM framework is to smooth the metric prior to linearization \cite{cohen2019certified}. Given $\sigma > 0$, $b_\sigma(\bt x) \doteq \E_{\mathbf{n} \sim \mathcal N(0, \sigma^2 \mathbf{I})} \big[ b(\bt x + \mathbf{n}) \big]$. Smoothing regularizes the metric by attenuating high-curvature components of its response surface, yielding a function with better local smoothness properties \cite{cohen2019certified}. Moreover, since
$\nabla b_\sigma(\bt x) = \E_{\mathbf{n}} \big[ \nabla b(\bt x + \mathbf{n}) \big]$,
the smoothed gradient is an average of perturbed gradients, favoring directions that remain stable under small perturbations of the input, which emphasizes features that may generalize better across multiple quality predictors. We use a Monte Carlo approximation with $n_s$ samples:
\begin{equation}
    b_\sigma(\img) = \sum_{i=1}^{n_s}\,  1/ n_s \, b(\img + \bt n_i), \quad \mathrm{with} \ \bt n_i \sim \mathcal{N}(\bt 0, \sigma\, \bt I),
\end{equation}
 and $\sigma$ is chosen such that the noise is imperceptible for a human observer \cite{smilkov2017smoothgrad}. Applying the same Taylor expansion around $\img$ as in \eqref{eq:Combined_NRMs}, we obtain a LNRM with the smoothed version of the gradient:
\begin{equation}
    \nabla b_\sigma(\img) = \sum_{i=1}^{n_s} 1/ n_s \, \nabla b(\img + \bt n_i).
\end{equation}
Then, $\nabla b_\sigma(\img)^\top(\cimg(\thetavec) - \img)$ is the smoothed LNRM (SLNRM). Since linearization decouples NRM evaluation from bit allocation, SLNRM still adds only limited complexity. However, the setup cost is higher for SLNRM: the initial LNRM requires only one backpropagation, while SLNRM requires $n_s$.

\subsection{LNRM with overfitted codecs}
LNRMs \eqref{eq:LNRM} can be applied to overfitted codecs. In this case, the main motivation is computational: optimizing the LNRM is much cheaper than optimizing the corresponding NRM. We focus on the complexity of a Cool-chic encoder with $n_i$ iterations. Let $n_{\sf f}$ be the number of FLOPs/px needed to compute a target NRM. Let $n_s$ be the number of smoothgrad samples. Since computing the gradient is twice as complex as evaluating the metric \cite{fernandez2025ratedist}, the complexity overhead due to the NRM is $3 \times n_i \times n_{\sf f}$ FLOPs/px, which increases to $3\times n_i \times n_{\sf f} \times n_s$ FLOPs/px when we consider smoothing. 

On the other hand, the complexity of evaluating the LNRM splits into two parts: computing the gradient, with complexity $3 \times n_{\sf f}$, and evaluating the LNRM on each iteration, with complexity $2 \, n_i$ FLOPs/px. For SLNRM, only the complexity of computing the gradient increases: we require $3 \times n_{\sf f} \times n_s$ FLOPs/px to obtain the smoothed gradient. Hence, relying on the LNRM reduces the computational complexity by a factor of approximately $n_i$. The number of iterations in Cool-chic is of the order of $10^4$; hence, the LNRM can reduce complexity by  orders of magnitude. \Cref{sec:exper} shows that complexity improvements also translate to runtime improvements. 

\begin{figure*}[!htbp]
\centering
\includegraphics[width=\linewidth]{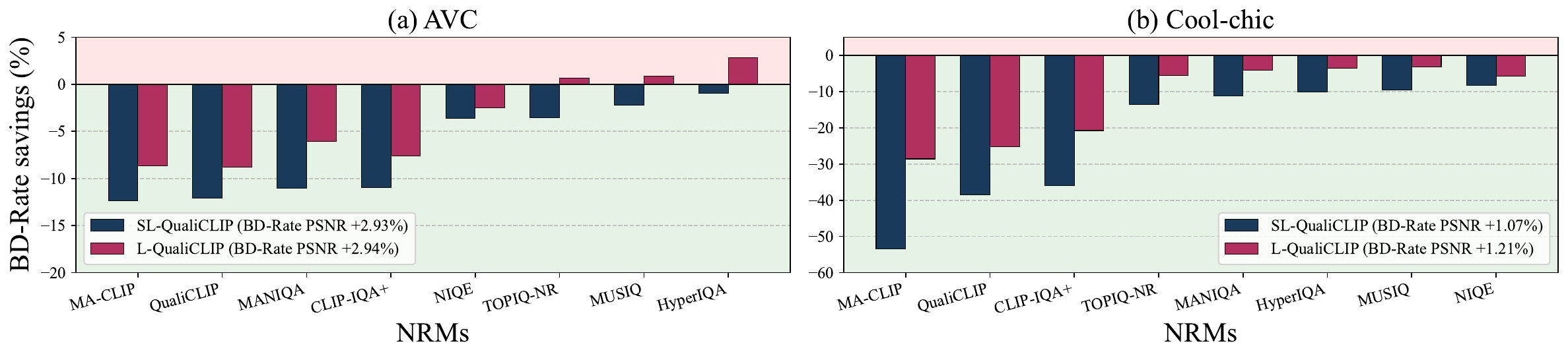}
\caption{BD-rate savings (\%) relative to the baseline across different NRMs. We optimize for the SLNRM (blue bar) and LNRM (red bar) versions of QualiCLIP. Left: Results for AVC. Right: Results for Cool-chic. Optimization with SLNRM yields higher BD-rate savings than LNRM across all NRMs we evaluated with. Moreover, using SLNRM always provides better scores than the baseline regardless of the NRM.
}
\label{fig:Exp_barchat_compare}
\end{figure*}

\begin{figure*}[ht]
    \centering
    
    \begin{subfigure}{\linewidth}
        \centering
        \includegraphics[width=\linewidth]{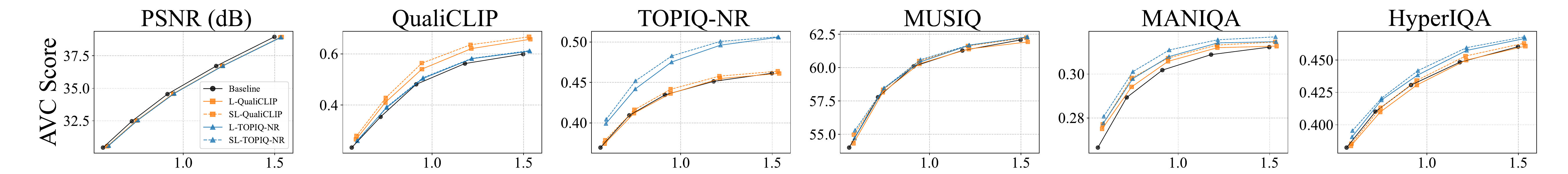}
    \end{subfigure}
    
    \begin{subfigure}{\linewidth}
        \centering
        \includegraphics[width=\linewidth]{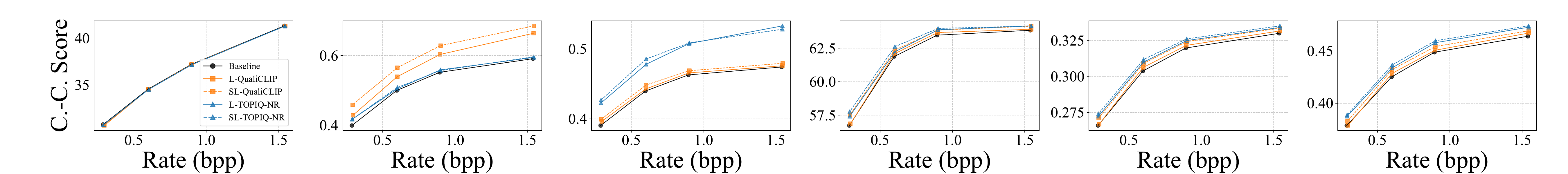}
    \end{subfigure}
    
    \caption{Rate-quality curves comparing baseline, LNRM, and SLNRM evaluated across various NRMs. 
    Top: the results for AVC; Bottom: the results for Cool-chic. In each sub-figure, the black curve represents the baseline. Other colors correspond to RDO using specific NRMs, where solid lines indicate LNRM and dashed lines indicate SLNRM. 
    SLNRM consistently outperforms LNRM across various NRMs.}
    \label{fig:RDcurves}
\end{figure*}

\begin{table*}[ht]
\centering
\caption{BD-rate savings (\%) relative to SSE-RDO AVC. Comparison of multiple LNRM and SLNRM pairs. SLNRM outperforms LNRM across various NRMs at comparable PSNR drops. Ensemble 1: QualiCLIP+CLIPIQA+, ensemble 2: QualiCLIP+ TOPIQ-NR.}
\label{tab:avc_bdrate}
\footnotesize
\setlength{\tabcolsep}{8pt}
\renewcommand{\arraystretch}{0.75}
\begin{tabularx}{\linewidth}{lccccccccc}
\toprule
Method & \textbf{PSNR} & \textbf{QualiCLIP} & \textbf{MA-CLIP} & \textbf{CLIP-IQA+} & \textbf{TOPIQ-NR} & \textbf{MANIQA} & \textbf{MUSIQ} & \textbf{HyperIQA} & \textbf{NIQE} \\
\midrule
\rowcolor{taborange!5} L-QualiCLIP  & 2.94 & -8.81 & -8.64 & -7.61 & 0.63 & -6.09 & 0.88 & 2.81 & -2.47 \\
\rowcolor{taborange!5} SL-QualiCLIP & 2.93 & -12.09 & -12.37 & -10.98 & -3.53 & -11.06 & -2.19 & -0.92 & -3.63 \\
\midrule
\rowcolor{tabblue!5} L-TOPIQ-NR & 3.28 & -2.51 & -2.37 & -4.07 & -22.56 & -10.85 & -1.85 & -6.25 & -2.77 \\
\rowcolor{tabblue!5} SL-TOPIQ-NR & 3.26 & -2.93 & -3.60 & -4.54 & -27.20 & -15.93 & -3.00 & -9.06 & -2.33 \\
\midrule
\rowcolor{taborange!5}  L-Ensemble 1 & 2.97 & -8.30 & -8.66 & -10.65 & 0.10 & -5.57 & 1.09 & 1.62 & -2.25 \\
\rowcolor{taborange!5}  SL-Ensemble 1 & 3.01 & -11.41 & -12.33 & -14.37 & -4.62 & -11.05 & -4.01 & -2.35 & -3.00 \\ \midrule
\rowcolor{tabpurple!5}  L-Ensemble 2 & 3.01 & -7.45 & -6.97 & -6.93 & -19.49 & -11.50 & -1.30 & -4.73 & -2.86 \\
\rowcolor{tabpurple!5}  SL-Ensemble 2 & 3.00 & -10.49 & -10.84 & -10.11 & -24.14 & -18.35 & -6.76 & -9.32 & -3.83 \\


\bottomrule
\end{tabularx}
\end{table*}


\begin{table*}[ht]
\centering
\caption{BD-rate savings (\%) relative to SSE-based Cool-chic. As with AVC, SLNRM consistently outperforms LNRM, and optimizing for the smoothed NRM outperforms optimizing for the NRM in the ensemble. When optimizing and evaluating with the same NRM, reconstructions may have no-overlapping RD curves with invalid BD-rates (dashed cases). We report the RD curves for these cases in \autoref{fig:rd_qua}.
}
\label{tab:cc_bdrate}
\footnotesize
\setlength{\tabcolsep}{8pt}
\renewcommand{\arraystretch}{0.75}  %
\begin{tabular*}{\linewidth}{l ccccccccc}
\toprule
Method & \textbf{PSNR} & \textbf{QualiCLIP} & \textbf{MA-CLIP} & \textbf{CLIP-IQA+} & \textbf{TOPIQ-NR} & \textbf{MANIQA} & \textbf{MUSIQ} & \textbf{HyperIQA} & \textbf{NIQE} \\
\midrule
\rowcolor{taborange!5} L-QualiCLIP  & 1.21 & -25.14 & -28.59 & -20.76 & -5.50 & -4.19 & -3.14 & -3.59 & -5.82 \\
\rowcolor{taborange!5} SL-QualiCLIP & 1.07 & -38.56 & -53.37 & -36.01 & -13.52 & -11.19 & -9.53 & -10.07 & -8.25 \\
\rowcolor{taborange!5} QualiCLIP & 0.82 & --- & --- & --- & -10.88 & -10.42 & -6.97 & -7.27 & -7.44 \\
\rowcolor{taborange!5} S-QualiCLIP & 0.75 & --- & --- & --- & -17.57 & -16.50 & -11.09 & -15.09 & -6.28 \\
\midrule
\rowcolor{tabblue!5} L-TOPIQ-NR & 0.95 & -5.66 & -3.61 & -5.41 & -43.07 & -13.34 & -7.25 & -14.56 & -3.52 \\
\rowcolor{tabblue!5} SL-TOPIQ-NR & 1.07 & -6.90 & -15.01 & -6.35 & -49.24 & -17.01 & -14.43 & -17.92 & -2.41 \\
\bottomrule
\end{tabular*}
\end{table*}

\subsection{SSE regularization}
\label{sec:regul}
As in \cite{fernandez2025ratedist}, we complement the LNRM with a SSE regularizer. To avoid modifying the Lagrangian, we constrain the SSE-based setup: 
\begin{multline}
\thetavec^\star = \argmin_{\thetavec\in\Theta} \ \norm{\cimg(\thetavec) - \img}_2^2  \  + \lambda \, \sum_{i = 1}^{n_b}\, r_i(\cimg_i(\theta_i)), \\
\mathrm{such \  that} \ \nabla \ell_c(\img)^\top(\cimg(\thetavec) - \img) \leq \mathrm{LNRM}_{\mathrm{max}},
\end{multline}
where $\mathrm{LNRM}_{\mathrm{max}}$ is the maximum acceptable LNRM value. By applying Lagrangian relaxation \cite{everett_generalized_1963}, we can write:
\begin{equation}
    d(\img, \cimg(\thetavec)) = \norm{\img - \cimg}_2^2 + \sum_{i = 1}^{n_m} \tau_i \, \nabla b_i(\img)^\top(\cimg - \img),
\end{equation}
where $\tau_i$ can be selected by the user to control the trade-off between performance among SSE and the different NRMs. To decide on the weights for each metric, we follow \cite{fernandez2025ratedist}: we split $\tau_i = \alpha \bar{\tau}_i$. Our choice of $\bar{\tau}$ varies depending on the type of codec. For hybrid codecs, we choose $\bar{\tau}_i =\sqrt{n_p/12} \,  \Delta / \norm{\nabla b_i(\img)}_2$, for all $i = 1, \hdots, n_m$, since it balances the SSE and LNRM contributions to the total cost for the smallest possible quantization error \cite{fernandez2025ratedist}. Overfitted codecs lack an explicit quantization step, so we cannot use the same relationship. Instead, we leverage the warm-up process to obtain estimates of both quantities. We have empirically found that the distortion term, whether SSE or NRMs, in overfitted codecs approaches its final value within a few iterations (\Cref{fig:warmup}). Thus, 
\begin{equation}
\label{eq:calibration}
    \bar{\tau}_i = \norm{\img - \cimg_{w}}_2^2 \, / \, b_i(\cimg
    _w),
\end{equation}
where $\cimg_w$ is the reconstruction at the end of the warm-up period. As a result, the user only needs to specify a single parameter $\alpha$, which controls the overall trade-off between the NRMs and the SSE. These parameters can be changed to align with user requirements.


\section{Empirical evaluation}
\label{sec:exper}
\noindent\textbf{Setup.} We evaluate two codecs: AVC \cite{wiegand_overview_2003} and Cool-chic \cite{ladune2023cool}. We test  50 frames of $640 \times 480$ pixels, sampled from the YouTube UGC dataset \cite{wang2019youtube}. We use 4:4:4 AVC baseline \cite{fernandez2025ratedist} with QP $\in\lbrace 25, 28, 31, 34, 37\rbrace$.  QP offset for color channels is $3$. We use RDO to choose block partition ($4\times 4$ and $16\times 16$) and block-level quantization step ($\Delta \mathrm{QP} = -4, -3, \hdots, 3, 4$). For computing $\bar{\tau}$ in \Cref{sec:regul}, we set $\Delta(\mathrm{QP}) = 2^{(\mathrm{QP} - 4)/6}$ \cite{ma2005study}. 
For Cool-chic, we use the fast training profile ($10^4$ iterations) with $\lambda$ values of 0.004, 0.001, 0.0004 and 0.0001. We use an AMD 9950X3D CPU with a NVIDIA Geforce RTX 5090 (32GB VRAM). The baseline is the version of each codec optimizing the SSE during RDO.

\begin{table}[ht]
\centering
\caption{BD-rate savings (\%) relative to SSE-RDO AVC as a function of smoothing parameters ($n_s$ and noise level $\sigma$). We optimize for SL-QualiCLIP. 
$n_s = 5$ and $\sigma = 0.01$ yields gains in all metrics.
}
\label{tab:exp_abs}
\footnotesize

\setlength{\tabcolsep}{1pt}
\renewcommand{\arraystretch}{1.0}  %
\begin{tabular}{lcccccc}
\toprule
$n_s,\sigma$ & {PSNR} & {QualiCLIP} & {CLIP-IQA+}  & {MUSIQ} & {MANIQA} & {HyperIQA} \\
\midrule
5,0.005 & 2.74 & -11.27 & -9.65 & 0.32 & -8.06 & 0.72 \\
5,0.01 & 2.93 & -12.09 & -10.98 & -2.19 & -11.06 & -0.92 \\
10,0.005 & 2.93 & -11.56 & -10.45 & -0.04 & -7.91 & 0.15 \\
10,0.01 & 2.89 & -12.72 & -11.52 & -3.13 & -12.72 & -2.12 \\
20,0.005 & 3.05 & -11.67 & -9.82 & -0.80 & -8.26 & -0.37 \\
20,0.01 & 2.91 & -12.73 & -11.47 & -3.35 & -11.85 & -3.26 \\
\bottomrule
\end{tabular}
\end{table}

\begin{table}[ht]
\centering
\caption{Runtime of gradient computation and Cool-chic encoding with LNRM, SLNRM, and NRM. Gradient computation time is reported in milliseconds (ms), while encoding time is shown as the overhead relative to the baseline ($\approx$ 50 s). Encoding with NRM introduces significant overhead compared to LNRM and SLNRM.}
\label{tab:exp_runtime}
\footnotesize
\setlength{\tabcolsep}{4pt}
\renewcommand{\arraystretch}{0.7}  %
\begin{tabular}{lccccc}
\toprule
 & \multicolumn{2}{c}{Gradient compute (ms)} & \multicolumn{3}{c}{Cool-chic encoding overhead (\%)} \\
\cmidrule(lr){2-3} \cmidrule(lr){4-6} 
NRMs & LNRM & SLNRM & LNRM & SLNRM & NRM \\
\midrule
\rowcolor{taborange!5}
QualiCLIP & 7 & 28 & +2.6\% & +2.8\% & +196\%\\
\rowcolor{tabblue!5}
TOPIQ-NR & 17 & 63 & +1.8\% & +2.6\% & +457\% \\
\bottomrule
\end{tabular}
\vspace{-2.0em}
\end{table}

We evaluate on QualiCLIP \cite{agnolucci2024quality}, TOPIQ-NR \cite{chen2024topiq}, CLIP-IQA+ \cite{wang2022exploring}, MANIQA \cite{yang2022maniqa}, MUSIQ \cite{ke2021musiq}, NIQE \cite{mittal2013making}, MA-CLIP \cite{liao2025beyond}, and HyperIQA \cite{Su_2020_CVPR} with default settings \cite{pyiqa}. For ensemble optimization, we analyze the NRM correlation (\autoref{fig:metrics_corr}) by assessing reconstruction quality using SSE-RDO AVC across 5 QP values. 
We identify three clusters; since one cluster only has one metric (NIQE), we select a NRM from the other two (TOPIQ-NR, QualiCLIP).

\begin{figure}[t]
    \centering
    \includegraphics[width=\linewidth]{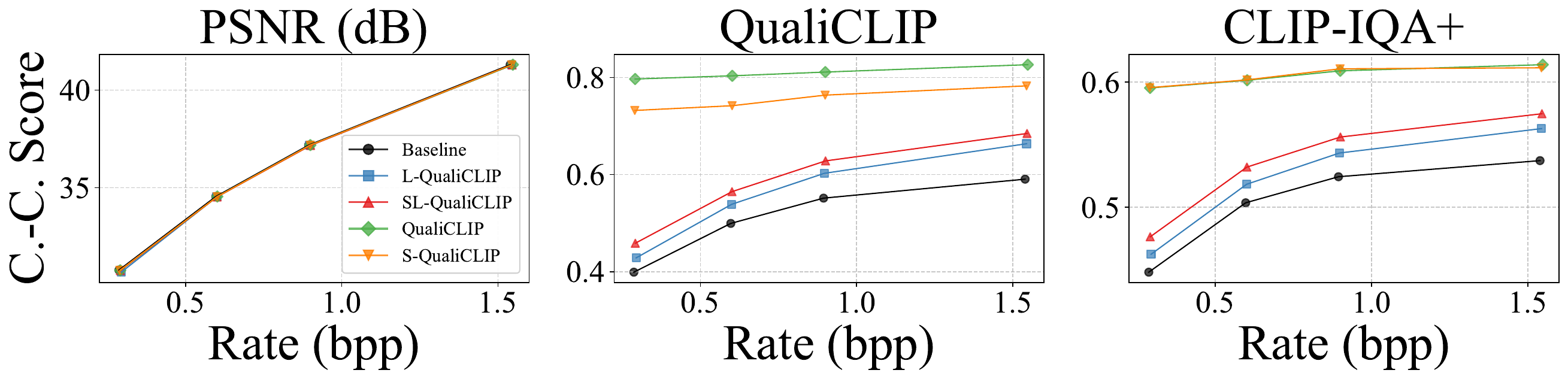}
    \caption{Rate-quality curves for Cool-chic. Using the NRM as a distortion metric in RDO performs best when evaluated by the same NRM. The SLNRM outperforms the LNRM in the given metric. }
    \label{fig:rd_qua}
\end{figure}

\noindent \textbf{Coding experiments.} We evaluate LNRM and SLNRM across multiple target NRMs. For smoothing, we set $n_s=5$ and $\sigma=0.01$. 
\autoref{fig:Exp_barchat_compare} illustrates the average BD-rate savings on various NRMs when optimizing for QualiCLIP via LNRM and SLNRM. 
We tune the hyperparameter $\alpha$ to make the BD-rate loss for PSNR comparable across all methods, implying that the degradation in SSE is similar across all methods. Under this condition, SLNRM consistently achieves higher bitrate savings than LNRM, not only for the target NRM but also across other evaluated metrics. 
We observe that the coding gains for Cool-chic are larger than those for AVC, which we attribute to the larger optimization space available on Cool-chic.
For AVC (\autoref{tab:avc_bdrate}), when optimizing for a single LNRM, performance is superior when evaluated on NRMs within the same cluster (cf.~\autoref{fig:metrics_corr}). When combining two target NRMs from the same cluster (e.g., QualiCLIP and CLIP-IQA+), we observe performance improvements on CLIP-IQA+ relative to using QualiCLIP only; however, gains on other metrics are negligible.
Conversely, when combining metrics from different clusters (e.g., QualiCLIP and TOPIQ-NR), we observe performance improvements in the TOPIQ-NR cluster relative to using QualiCLIP alone. Applying smoothing to these combinations also yields broad improvements across all evaluated metrics. 
For Cool-chic, we extend our analysis to include direct optimization of both the NRM and its smoothed version. As shown in \autoref{fig:rd_qua}, when optimizing and evaluating for QualiCLIP, directly optimizing the NRM yields the best performance, whereas the smoothed NRM shows a slight degradation.  Yet, when evaluating NRMs from different clusters (\autoref{tab:cc_bdrate}), the smoothed NRM outperforms direct NRM optimization.
These results suggest that optimizing the smoothed metric directly also improves performance in ensembles of NRMs.

\noindent\textbf{Ablations} (\autoref{tab:exp_abs}). We test smoothing parameters with QualiCLIP. We observe that $n_s=5$ provides a good trade-off between performance and computational overhead. Increasing $n_s$ yields marginal gains, with performance saturation beyond $n_s = 10$. Since the computational cost of smoothing scales linearly with $n_s$, we adopt $n_s = 5$ for our experiments. We also evaluate the impact of noise intensity with $\sigma=0.005$ and $\sigma=0.01$. The visual impact of the added noise is negligible, with PSNR values relative to the input UGC image of $46.1$ dB and $40.1$ dB, respectively. A larger $\sigma$ ($0.01$) leads to better coding gains across all the evaluated NRMs.

\noindent \textbf{Runtime analysis} (\autoref{tab:exp_runtime}).
We show the runtime for gradient computation and Cool-chic encoding with LNRM, SLNRM, and NRM for five images with two $\lambda$ values. 
The experiment is repeated twice with two random seeds.
Though computing smoothed gradients with $n_s=5$ increases complexity by approximately $5$ times, this overhead is negligible relative to the Cool-chic encoding time (around $50$ seconds per image in our setup). Hence, the difference in encoding runtime between LNRM and SLNRM is minimal. Directly optimizing for NRM incurs a substantial runtime penalty, since the NRM must be evaluated at each iteration.

\section{Conclusion}
In this paper, we showed that optimizing for a single NRM leads to inconsistent behavior when evaluated with other NRMs. To address this, we proposed an RDO framework based on LNRMs that enables optimization over ensembles of metrics. Building on this formulation, we introduced a smoothing-based approach inspired by robust optimization, which improves performance not only on the target metric but also in other NRMs while exploiting the computational advantages of linearization. Experiments with  AVC and Cool-chic on the YouTube UGC dataset demonstrate consistent bitrate savings across diverse NRMs, reduced sensitivity to the choice of NRM, and substantial encoder runtime reductions compared to direct NRM optimization for Cool-chic, with no decoder-side overhead. While we focused on image compression, the proposed framework naturally extends to video coding by incorporating video NRMs.
\vfill\pagebreak
\clearpage


\bibliographystyle{IEEEbib}
\bibliography{refs}

\end{document}